\documentclass[iop,twocolappendix]{emulateapj}
\usepackage{amsmath,amstext}
\usepackage{bm}
\usepackage{mathtools}
\usepackage[T1]{fontenc}
\usepackage[normalem]{ulem}
\usepackage{hyperref}
\usepackage[figure,figure*]{hypcap}
\hypersetup{colorlinks=true,citecolor=blue,linkcolor=blue,urlcolor=blue}

\newcommand{\cta}[1]{\citetalias{#1}}

\DeclareMathOperator*{\argmin}{argmin}
\DeclareMathOperator*{\erf}{erf}

\defcitealias{M07}{M07}
\defcitealias{J12}{J12}
\defcitealias{C12}{C12}
\defcitealias{J14}{J14}

\begin{document}

\title{Sparse reconstruction of the merging A520 cluster system}

\author{Austin Peel\altaffilmark{1}}
\affil{D{\'e}partement d'Astrophysique, IRFU, CEA, Universit{\'e} Paris-Saclay, F-91191 Gif-sur-Yvette, France}
\email{austin.peel@cea.fr}

\author{Fran\c{c}ois Lanusse}
\affil{McWilliams Center for Cosmology, Department of Physics, Carnegie Mellon University, 
             Pittsburgh, PA 15213, USA}
             
\and

\author{Jean-Luc Starck\altaffilmark{1}}
\affil{D{\'e}partement d'Astrophysique, IRFU, CEA, Universit{\'e} Paris-Saclay, F-91191 Gif-sur-Yvette, France}
             
\altaffiltext{1}{Universit{\'e} Paris Diderot, AIM, Sorbonne Paris Cit{\'e}, CEA, CNRS, F-91191 Gif-sur-Yvette, France}

\begin{abstract}
Merging galaxy clusters present a unique opportunity to study the properties of 
dark matter in an astrophysical context. These are rare and extreme cosmic events in which the 
bulk of the baryonic matter becomes displaced from the dark matter halos of the colliding subclusters. 
Since all mass bends light, weak gravitational lensing is a primary tool to study the total mass 
distribution in such systems. Combined with X-ray and optical analyses, mass maps of cluster mergers 
reconstructed from weak-lensing observations have been used to constrain the self-interaction 
cross-section of dark matter. The dynamically complex Abell 520 (A520) cluster is an exceptional case, 
even among merging systems: multi-wavelength observations have revealed a surprising high mass-to-light 
concentration of dark mass, the interpretation of which is difficult under the standard assumption of 
effectively collisionless dark matter. We revisit A520 using a new sparsity-based mass-mapping algorithm 
to independently assess the presence of the puzzling dark core. We obtain high-resolution mass 
reconstructions from two separate galaxy shape catalogs derived from Hubble Space Telescope observations 
of the system. Our mass maps agree well overall with the results of previous studies, but we find 
important differences. In particular, although we are able to identify the dark core at a certain level 
in both data sets, it is at much lower significance than has been reported before using the same data. 
As we cannot confirm the detection in our analysis, we do not consider A520 as posing a significant 
challenge to the collisionless dark matter scenario.
\end{abstract}

\keywords{dark matter -- galaxies: clusters: individual (Abell 520) -- gravitational lensing: weak}

\section{Introduction}\label{sec:intro}
As the largest gravitationally bound objects in the Universe, galaxy clusters represent the most
recent phase in the hierarchical formation of cosmic structure. Understanding their properties 
and evolution is crucial to the progress of cosmology. Multi-wavelength studies
have revealed that the majority of the mass in clusters resides in cold dark matter (CDM), while the
bulk of the baryonic matter is contained in the hot intracluster gas. The galaxies themselves
constitute only a few percent of the total mass. The often extreme characteristics of merging 
cluster systems make them a unique astrophysical laboratory in which to test the paradigm of 
collisionless dark matter. 

The primary signature of a merger between two (or more) clusters is the dissociation
of the intracluster gas from the dark matter and the galaxies, the latter two of which remain 
spatially coincident. 
This is because the galaxies interact principally via the tidal gravitational fields and thus 
essentially pass through each other. In contrast, the ionized intracluster plasma clouds 
experience ram pressure that slows them down during crossing, leaving an overdensity of X-ray 
emitting gas between the luminous subclusters along the merger axis. The fact that the dark
mass component, inferred by, for example, weak-lensing analysis, remains separate from the bulk 
of the baryons has been seen as direct proof of the existence of dark matter 
\citep{CGM.2004, CBG.etal.2006}.
Furthermore, the relative positions of the dark matter and galaxy centroids have led to
an upper limit on the the self-interaction cross-section of dark matter; see, for example,
\citet{MGC.etal.2004, RMC.etal.2008, Bradac2008, KSHF.etal.2014, Harvey2015, RME.2017}. 

The Abell 520 system (MS 0451+02, $z=0.2$, \citet{Abell1989}), first studied using weak lensing by 
\citet[hereafter M07]{M07}, exhibits complex structure and
offers a possible counterexample to the collisionless dark matter scenario. 
Like the Bullet Cluster (1E 0657-558) \citep{MGL.etal.2002}, the galaxies and dark matter in A520 are 
offset from the intracluster gas distribution, indicating that significant ram-pressure stripping has 
occurred from merging. 
It is our fortunate viewing angle with respect to the orientation of the merger axis, which lies 
essentially in the plane of the sky, that allows us to observe these offsets.
A recent X-ray study using deep
\textit{Chandra}\footnote{\url{http://chandra.harvard.edu/}} data has elucidated details of the history 
of the merger using structure-rich temperature maps \citep{WMG.2016}.
However, the authors of \cta{M07} also detected a dark core, labeled as P3 in the figures below,
using data from the Canada-France-Hawaii Telescope\footnote{\url{http://www.cfht.hawaii.edu/}} (CFHT) 
and Subaru\footnote{\url{http://subarutelescope.org/}}. 
The unexpected dark structure was found to coincide 
spatially with the peak of the X-ray emission, but unlike the other detected mass peaks, the dark 
core region did not appear to contain any luminous cluster galaxies. The presence of such a high 
mass-to-light-ratio peak challenges the current understanding of dark matter and, if real, could 
help to significantly reduce the parameter space of possible dark matter particle candidates.

\citet{OU.2008} performed a reanalysis of the data of \cta{M07} and found a 
consistent mass peak at the dark core location in their reconstruction.
Two further follow-up studies of A520 were carried out in 2012 with weak-lensing analyses 
presented in \citet[hereafter J12]{J12} and \citet[hereafter C12]{C12}. 
\cta{J12} confirmed the dark peak detection at higher than $10\sigma$ 
significance using mosaic images in a single passband from the
\textit{Hubble Space Telescope} (\textit{HST}) Wide Field Planetary Camera
2\footnote{\url{http://www.stsci.edu/hst/wfpc2}} (WFPC2). 
The higher resolution
data afforded a number density of galaxies more than 3 times higher than CFHT. The reconstructed 
mass map of \cta{J12} agreed overall with that of \cta{M07}, both in the 
positions of the most significant mass peaks, as well as their aperture mass estimates.
Two new peaks were detected in \cta{J12}, labeled as P5 and P6, the former of which resolved a 
discrepancy from 
\cta{M07}, where the peak was expected but curiously absent from the previous reconstruction.
The other new peak emerged due to the higher resolution of the data at a location consistent with
some of the bright cluster members.

\begin{figure*}[t]
\centering
\includegraphics[width=17.8cm]{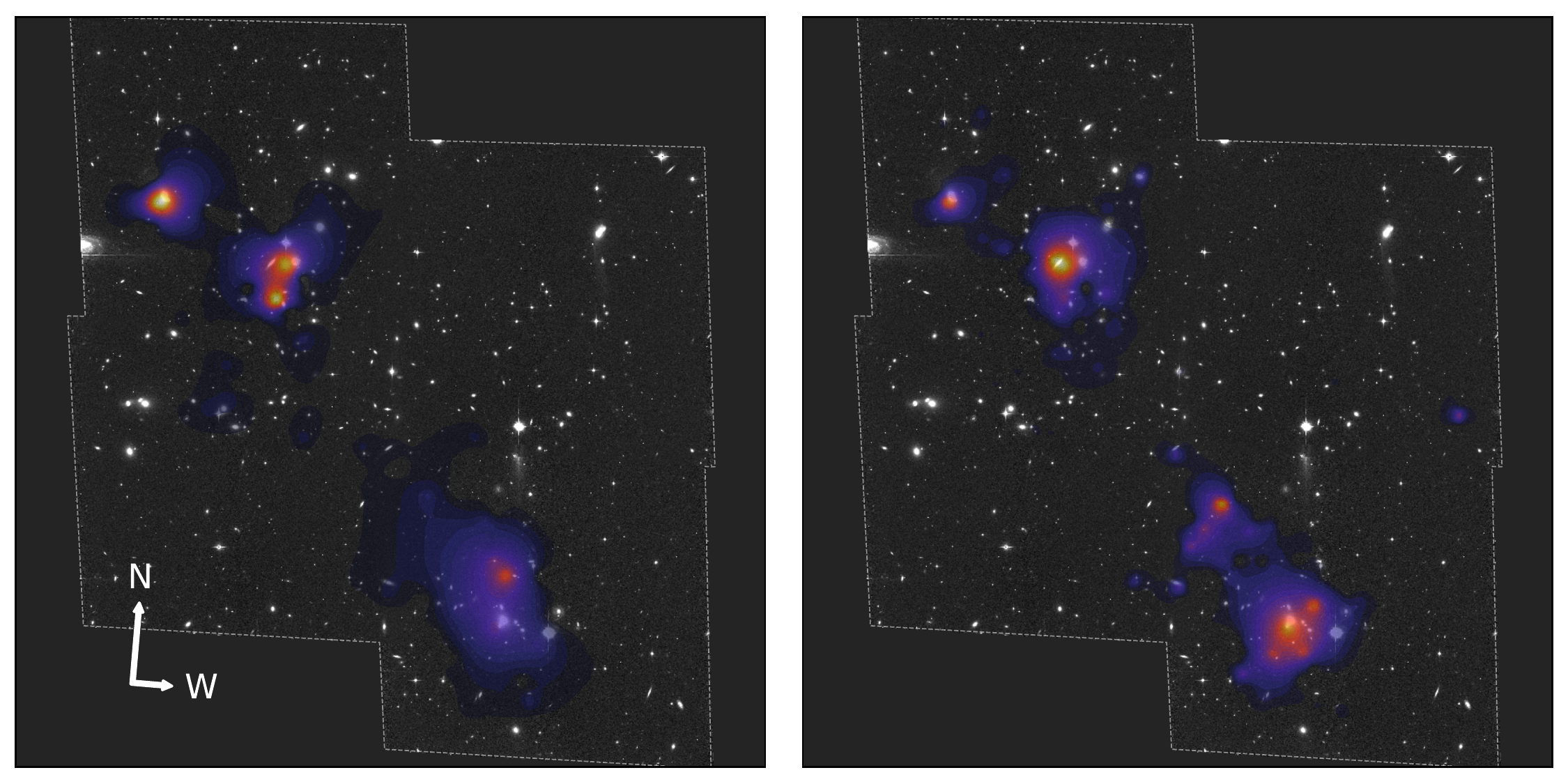}
\caption{Example Glimpse2D mass map reconstructions using the weak-lensing catalogs of \cta{C12} 
(left) and \cta{J14} (right) overlaid on the \cta{J14} science image.}
\label{fig:sci_img}
\end{figure*}

Contrary to the three previous studies, \cta{C12} did not detect a dark core in 
their data, which consisted of independent ground-based 
Magellan\footnote{\url{http://www.lco.cl/telescopes-information/magellan/}} observations combined 
with mosaic images from the \textit{HST} Advanced Camera for 
Surveys\footnote{\url{http://www.stsci.edu/hst/acs/}} (ACS).
The locations, morphologies, and aperture mass 
measurements of the primary cluster substructures are mostly consistent with those of 
\cta{J12}, although a few significant differences were found. These include
larger uncertainties in mass estimates for all peaks, as well as measuring a larger luminosity 
and a lower column mass at the dark peak location. \cta{C12} concluded that
the gross mass distribution of A520 is consistent with a constant mass-to-light ratio and that 
both \cta{M07} and \cta{J12} overstated the significance of 
their dark peak detections.

More recently, \citet[hereafter J14]{J14} revisited A520 with an updated weak-lensing
analysis using ACS data and also provided a detailed comparison with \cta{J12} 
and \cta{C12}. 
The study claimed again to find dark peak region characterized by a very high mass-to-light ratio, 
although not at the same location as in \cta{M07} and \cta{J12}.
The position of the new peak P3$'$ was offset from the former by about 1 arcmin southwest toward the
largest mass substructure P4. In contrast to \cta{C12}, a $\chi^2$ test 
led the authors to reject the constant mass-to-light ratio hypothesis at a level of at least $\sim6\sigma$.
Comparing catalogs and mass reconstructions with \cta{C12} indicated that
the discrepancies were likely caused by differences in the charge transfer inefficiency (CTI) 
correction methods and the shape measurement pipelines. It is not clear what the origin of a true 
dark peak in the A520 data would be, although a number of scenarios were suggested in \cta{M07}, 
\cta{J12}, and \cta{J14}. The most intriguing possibility is that dark matter particles could 
possess a non-negligible self-interaction cross-section.

Given the disagreement in the literature and the scientific impact that detecting a real dark 
substructure would have, we perform new mass map reconstructions of A520 using a completely 
different algorithm from those of the previous studies. The software 
is called Glimpse2D \citep{LSL.etal.2016}, and it approaches the mass-mapping problem as an
ill-posed inverse problem, regularized by a multi-scale wavelet prior on the reconstructed
surface mass density map. The algorithm is able to retain all available small-scale information
by avoiding the need to bin the irregularly sampled shear field. It has also been shown to 
perform beautifully on ACS-like weak-lensing simulations, reproducing the input maps at high resolution
and fidelity. The goal of this work is therefore twofold: the first is to test Glimpse2D on real 
data, and the second is to determine whether we detect a significant dark peak in accordance 
with \cta{J14}.

The remainder of the paper is organized as follows. In Section \ref{sec:data} we describe the two 
galaxy catalogs we use in our weak-lensing analysis, which correspond to those of \cta{C12} and \cta{J14}. 
We review some basics of weak-lensing theory and describe our sparsity-based approach to mass mapping in 
Section \ref{sec:method}. In Section \ref{sec:results} we present our mass map reconstructions of A520,
along with uncertainty and significance analyses. 
We summarize and conclude in Section \ref{sec:conclusion}.

\section{Data}\label{sec:data}
We obtained both weak-lensing catalogs used in \cta{C12} and \cta{J14} (private communication) 
to carry out our sparsity-based surface mass reconstructions of A520. We give brief 
descriptions of these data sets here and refer to their respective papers for more complete details.
As an illustration of the data field and its primary features, we show mass map contours
derived from Glimpse2D using typical parameters overlaid on the science image of \cta{J14} 
in Figure \ref{fig:sci_img} (compare Figure 5 of their paper). Contours on the left and right 
plots were obtained using the \cta{C12} and \cta{J14} data, respectively. Details of the algorithm 
and analysis results are presented fully in the following sections.

\subsection{C12: Magellan + HST/ACS}\label{ssec:C12data}
The catalog used for the weak-lensing analysis in \cta{C12} was derived from a 
combination of images from the Magellan telescope along with an \textit{HST}/ACS mosaic of 
four pointings (PI: D. Clowe).
Magellan imaged A520 in Bessel B, V, and R passbands with a field of view of 15.4 arcmin. These 
images were cleaned of defects, corrected for smearing by the point-spread function (PSF), 
and co-added to produce a final image. 
The ACS data cover a smaller area near the center of the Magellan field and consist of four
partially overlapping fields each imaged in the filters F435W, F606W, and F814W. An important 
step in reducing these data was correcting for the effect of charge transfer inefficiency (CTI)
caused by the degradation of the CCD detectors due to prolonged radiation exposure of the 
instrument outside Earth's atmosphere. CTI induces spurious distortions in the shapes of galaxies 
in a way that can substantially contaminate the weak-lensing signal. \cta{C12} obtained 
consistent results from two independent CTI correction procedures and concluded that CTI did not 
significantly impact their final weak-lensing analysis.

\begin{figure*}
\centering
\includegraphics[width=18cm]{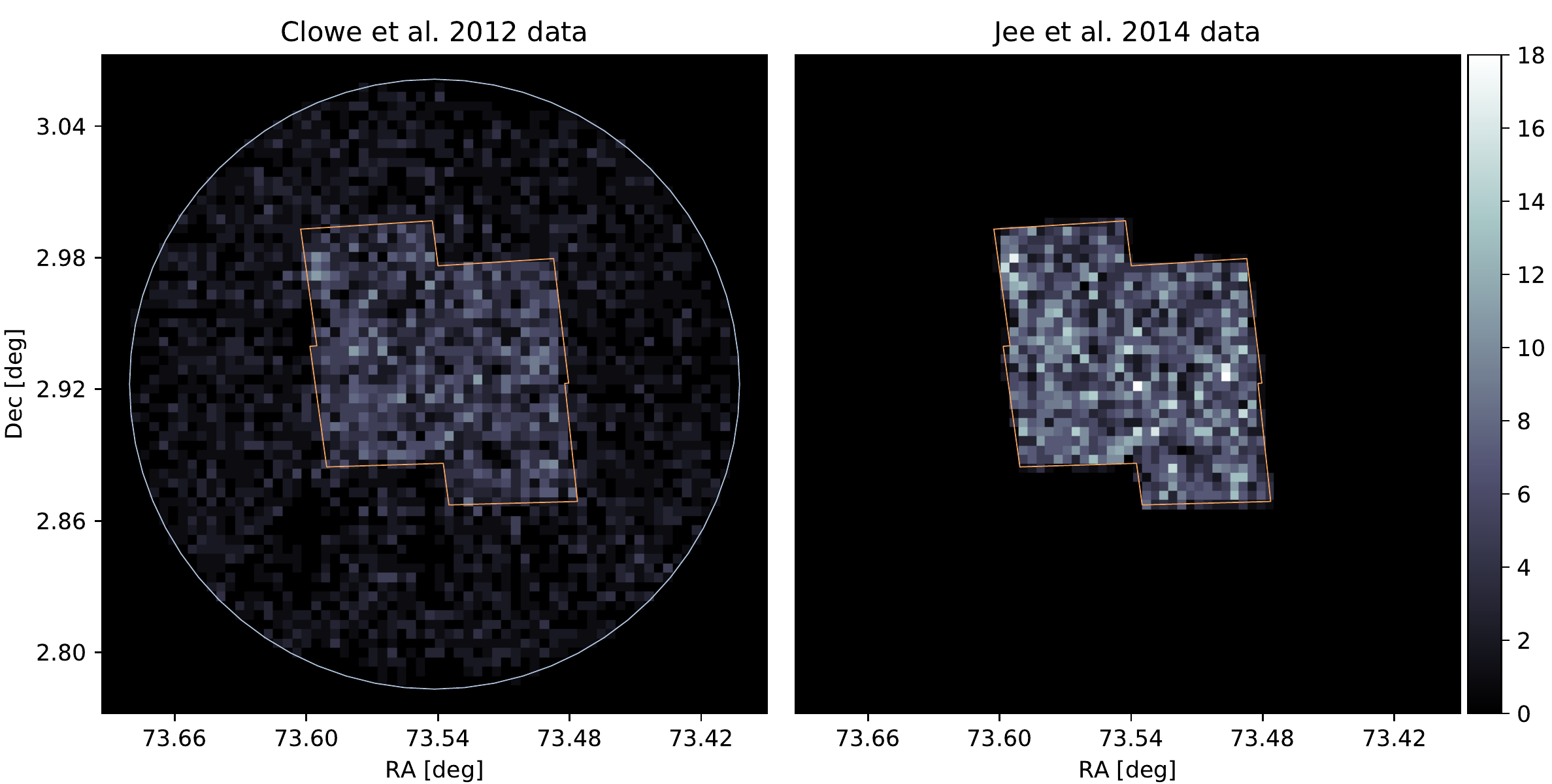}
\caption{Source galaxy density maps from \cta{C12} (left) and \cta{J14} (right) 
         data sets. The \cta{C12} ellipticity catalog is derived from a combination of Magellan 
         and \textit{HST}/ACS images with galaxy number densities of 22 arcmin${}^{-2}$ and 56 
         arcmin${}^{-2}$, respectively. The \cta{J14} catalog comes from the same ACS data 
         as \cta{C12}, but exhibits a higher galaxy density of 109 arcmin${}^{-2}$ due to a 
         different reduction pipeline.}
\label{fig:galaxydensity}
\end{figure*}

Shape measurements were made separately on galaxies in the Magellan and ACS images and then 
combined for a total of 5903 objects in the final lensing catalog. To be compatible with the ACS
galaxies, the Magellan set of observed ellipticities was scaled by a small factor to account for 
the difference in the redshift distributions between the two populations.
A map of the source galaxy density 
is shown the left plot of Figure \ref{fig:galaxydensity}. The outer circle marks the boundary 
of the Magellan field, and the inner polygonal area indicates the ACS footprint. The central ACS 
region, which contains the majority of the cluster mass, shows a higher source density than its 
surroundings, as expected from space-based observations. 

\subsection{J14: CFHT + HST/ACS}\label{ssec:J14data}
The \cta{J14} catalog was derived from the same raw ACS images that were used in 
\cta{C12} combined with the ground-based CFHT catalog that was used in \cta{J12}.
Different reduction procedures of the ACS images led to a different set of galaxies in
the final catalog. Notably, the \cta{J14} catalog contains approximately twice the number 
of galaxies as that of \cta{C12}, the difference arising partly due to the 
inclusion of more faint galaxies by \cta{J14}. One reason for this could stem from the
different drizzling kernels used to create the mosaics. \cta{J14} used an approximate
sinc interpolation kernel, whereas \cta{C12} used a square kernel that may not perform as
well in measuring the shapes of small galaxies. Perhaps more importantly, the CTI correction 
method used by \cta{J14} was updated with improved performance in the low-flux regime
\citep{UA.2012} compared to what was available to \cta{C12}. A comparison between the 
catalogs of \cta{C12} and \cta{J14} revealed no significant difference between the
the (independently) calibrated ellipticity components for their common galaxies \citep{J14}. 

\cta{J14} used the F814W image for their primary WL analysis, although the other filters
were used for identifying and removing foreground/cluster galaxies. It is worth noting that 
\cta{J14} claim that although both studies supplement their ACS data with (different) 
ground-based observations, the difference between their mass reconstructions comes from the 
treatment of the ACS data.

The \cta{J14} source galaxy density is shown in the right panel of Figure \ref{fig:galaxydensity}
using the same color scale as the left panel to indicate pixel number counts.
The map is visually consistent with the ACS region of \cta{C12}, but the higher number density 
of the \cta{J14} catalog of 4953 galaxies is clearly seen. We note that the pixelization in
Figure \ref{fig:galaxydensity} has been chosen simply for visualization purposes; we use a much 
higher resolution in our mass reconstructions.

\section{Method}\label{sec:method}
Different mass-mapping techniques were used in the \cta{C12} and \cta{J14} analyses. \cta{C12}
used an improved version of Kaiser-Squires inversion \citep{KS.1993} that accounts for the 
reduced shear \citep{SS.1995}. On the other hand, \cta{J14} used an implementation 
\citep{JFI.etal.2007} of the entropy-regularized maximum likelihood method introduced by 
\citet{SSB.1998}. Our mass reconstruction technique is completely independent of these two methods.
We present a summary of Glimpse2D in this section after briefly recalling some basics of 
weak-lensing theory. 

\subsection{Weak-lensing and mass maps}
Galaxy shape distortions caused by gravitational lensing can be characterized by a transformation
between the lensed image coordinates $\bm\theta$ and the source coordinates $\bm\beta$.
In the linear regime, the mapping between $\bm\theta$ and $\bm\beta$ is given by the amplification
matrix $\mathcal{A}=\partial\bm\beta/\partial\bm\theta$, which is parameterized by a scalar part 
$\kappa(\bm\theta)$ and a complex (spin-two) field $\gamma(\bm\theta)$ as
\begin{equation}
   \mathcal{A} = 
     \begin{pmatrix}
       1 - \kappa - \gamma_1 & -\gamma_2 \\
       -\gamma_2 & 1 - \kappa + \gamma_1
     \end{pmatrix}.
\end{equation}

The function $\kappa$ is called \textit{convergence} and quantifies an isotropic change in the 
size of the source image, while the \textit{shear} $\gamma$ describes anisotropic stretching. 
In the context of lensing by large-scale structures, 
both $\kappa$ and $\gamma$ are much smaller than 1. The convergence can also be interpreted 
directly as the projected mass density of the matter field between the observer and the source.
As such, it is often convenient to express $\kappa$ as
\begin{equation}
  \kappa = \frac{\Sigma}{\Sigma_\mathrm{crit}},
\end{equation}
where $\Sigma(\bm\theta)$ is the surface mass density of the lens, and $\Sigma_\mathrm{crit}$ is 
the critical surface mass density given by
\begin{equation}
  \Sigma_\mathrm{crit} = \frac{c^2}{4\pi G}\frac{D_\mathrm{S}}{D_\mathrm{L}D_\mathrm{LS}}.
\end{equation}
In the above equation, $D_\mathrm{S}$, $D_\mathrm{L}$, and $D_\mathrm{LS}$ are the angular 
diameter distances from observer to source, observer to lens, and from lens to source, 
respectively.

The convergence and shear are both expressible as derivatives of a scalar lensing potential 
$\psi$. This leads to an integral relation between $\kappa$ and $\gamma$ \citep{KS.1993},
\begin{equation}
  \kappa(\bm\theta) - \kappa_0 = \frac{1}{\pi}\int \mathrm{d}^2\theta' 
                                 \mathcal{D}^*(\bm\theta - \bm\theta') \gamma(\bm\theta'),
\label{eq:kappa_from_gamma}
\end{equation}
where $\kappa_0$ is a constant of integration corresponding to the mass-sheet degeneracy, and
the kernel $\mathcal{D}$ is given by
\begin{equation}
  \mathcal{D}(\bm\theta) \coloneqq -\frac{\theta_1^2 - \theta_2^2 + 2\mathrm{i}\theta_1\theta_2}{|\bm\theta|^4}.
\end{equation}

The reconstruction of a convergence map is hindered in practice by several considerations. First, 
the shear field is not directly observable, lensing surveys measure instead the reduced shear 
$g=\gamma/(1-\kappa)$. Second, Equation (\ref{eq:kappa_from_gamma}) assumes knowledge of the 
shear field over an infinite domain, in practice the reduced shear is only sampled at discrete 
points over a limited survey area. Finally, in most situations, the shear signal is dominated 
by shape noise, some sort of filtering is therefore required to recover a meaningful map. In this 
work, we reconstruct the convergence map using the Glimpse2D algorithm \citep{LSL.etal.2016} which 
aims at addressing all three issues simultaneously.

\subsection{Glimpse2D: sparsity-based mass mapping}\label{ssec:glimpse}
For the reasons stated in the previous section, mass-mapping is in practice a non-trivial inverse 
problem. Glimpse2D aims at solving  this inverse problem using sparse recovery, a powerful 
framework for solving ill-posed inverse problems with many successful applications in image 
processing, medical imaging, radio-interferometry and astrophysics. This approach relies on 
the so-called sparse prior, the idea that when expressed in an appropriate basis, or dictionary, 
most signals are sparse (i.e. have only a few non-zero coefficients) or at least compressible 
(i.e. can be represented  to a very good approximation by a sparse signal). More formally, a 
signal $\bm{x}$ is said to be sparse in a dictionary $\mathbf{\Phi}$ if only a small number 
of its coefficients $\bm{\alpha} = \mathbf{\Phi}^* \bm{x}$  are non-zero. Consider a 
generic linear inverse problem of the form:
 \begin{equation}
  \bm{y} = \bm{A}\bm{x} + \bm{n},
\label{eq:gen_lin_inverse}
\end{equation}
where $\bm{y}$ are the measurements, $\bm{A}$ is a linear operator, $\bm{x}$ is the signal
we want to recover, and $\bm{n}$ is an additive noise term.  Under the assumption that  the 
signal to recover $\bm{x}$ is sparse in dictionary $\mathbf{\Phi}$, the solution can be 
robustly estimated by solving an optimization problem  of the form:
\begin{equation}
  \overline{\bm{x}} = \argmin_{\bm{x}}\;\frac{1}{2} \lVert\bm{y} - \bm{A}\bm{x}\rVert_2^2 + 
                 \lambda \lVert\bm{\Phi}^*\bm{x}\rVert_1\,
\label{eq:convex_opt}
\end{equation}
where $\lambda$ is a regularization parameter. The objective function in Eq. (\ref{eq:convex_opt}) 
seeks a solution that balances data fidelity, via the first (quadratic) term, against the sparsity 
of the analysis coefficients $\bm\alpha=\bm{\Phi}^*\bm{x}$ via the sparsity inducing $\ell_1$ 
norm of the second term. 

\begin{figure*}
\centering
\includegraphics[width=18cm]{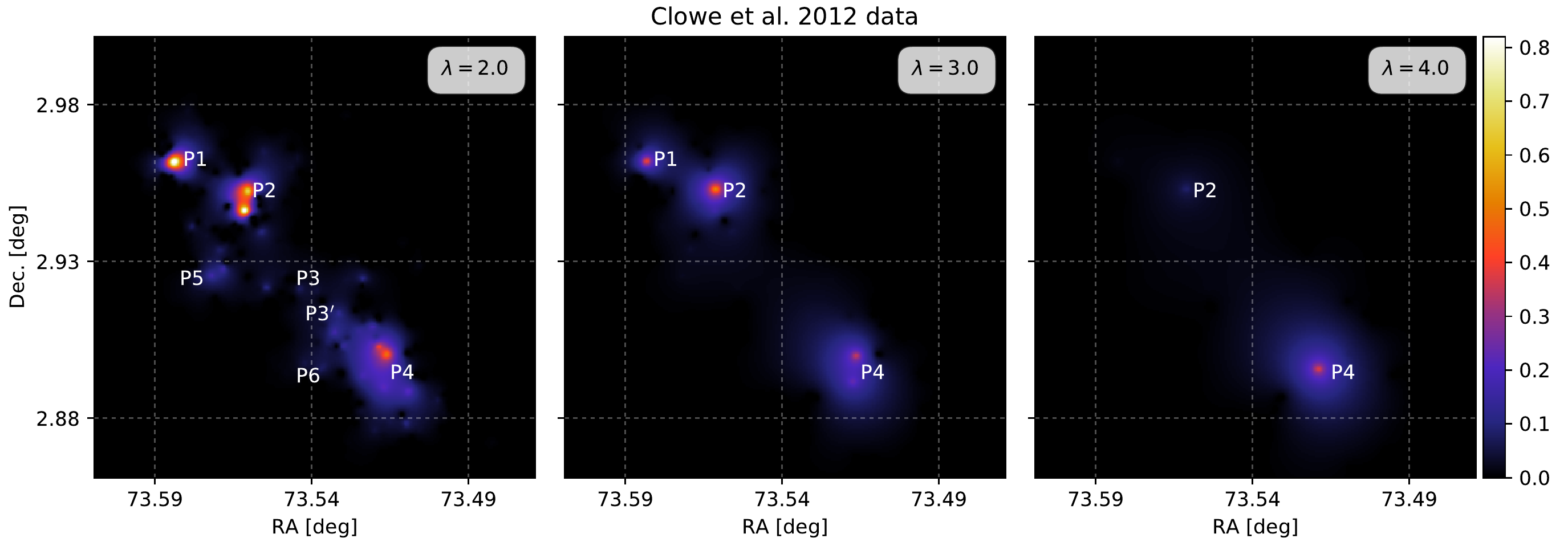}
\caption{Surface mass reconstructions from \cta{C12} data for regularization parameters $\lambda=2.0$, 
         3.0, and 4.0. Labels P1--P6 indicate the approximate locations of the relevant structures reported in
         \cta{C12} and \cta{J14}. From left to right, one can see that noise and low-amplitude features are 
         better suppressed with increasing $\lambda$. The presence of a dark core at P$3'$ claimed by 
         \cta{J14} is visible in the $\lambda=2.0$ map, but not in the $\lambda=3.0, 4.0$ maps.}
\label{fig:fig1}
\end{figure*}

Glimpse2D adopts this framework to treat the full mass-mapping problem, from noisy and discretely 
sampled noisy reduced shear measurements to a non parametric mass-map. In this situation, the 
inverse problem that we aim to solve takes the following form:
\begin{equation}
	\bm{g} = \frac{\bm{Z} \mathbf{T} \mathbf{P} \mathbf{F}^* \bm{\kappa}}{1 - \bm{Z} \mathbf{T} \mathbf{F}^* \bm{\kappa} } + \bm{n} \;,
	\label{eq:inv_prblm}
\end{equation}
where $\mathbf{F}$ is the Discrete Fourier operator, $\mathbf{P}$ is the Fourier-based lensing 
operator, yielding $\hat{\bm{\gamma}}$ from $\hat{\bm{\kappa}}$, $\mathbf{T}$ is the 
Non-equispaced Discrete Fourier-Transform (NDFT) operator evaluating the Fourier-Transform at 
the positions of the galaxies in the catalog, and $\mathbf{Z}$ is a cosmological weight 
function depending on the redshifts of the lens and background sources. Here $\bm{\kappa}$ is 
understood to be the convergence for sources at infinite redshift and $\mathbf{Z}$ rescales 
the convergence based on the redshift of each individual galaxy in the survey. We refer to 
\cite{LSL.etal.2016} for more details on the definition of these operators. 
Equation~\eqref{eq:inv_prblm} relates the convergence map $\bm{\kappa}$  to the reduced shear 
$\bm{g}$ measured at each galaxy position, Glimpse2D aims at recovering $\bm{\kappa}$ by 
solving the following optimization problem:
\begin{multline}
 \argmin_{\bm{\kappa}} \frac{1}{2} \parallel \mathcal{C}_\kappa^{-1} \left[ (1 - \bm{Z} \mathbf{T} \mathbf{F}^* \bm{\kappa}) \bm{g} - \bm{Z}\mathbf{T} \mathbf{P} \mathbf{F}^* \bm{\kappa} \right] \parallel_2^2 \\ + \lambda \parallel \bm{w} \circ \bm{\Phi}^* \bm{\kappa} \parallel_1 + i_{\Im(\cdot) = 0}(\bm{\kappa}) \;.
 \label{eq:algo_rec_non_lin_linearised_z}
\end{multline}
where the matrix 
$\mathcal{C}_\kappa^{-1} = \Sigma^{-\frac{1}{2}}/(1 -\bm{Z}  \mathbf{T} \mathbf{F}^* \bm{\kappa})$ 
accounts for a diagonal covariance matrix of the lensing measurements $\Sigma$, $\lambda$ is our 
regularization parameter, $\bm{w}$ is a sparsity scaling factor based on the level of noise in 
the data, and  $ i_{\Im(\cdot) = 0}()$ is an indicator function ensuring that the reconstructed 
convergence map has no imaginary part, i.e. enforcing a  zero B-mode constraint. Again, we refer 
the  reader to \cite{LSL.etal.2016} for the details of the algorithm\footnote{The Glimpse2D
software is publicly available at \url{http://www.cosmostat.org/software/glimpse}} solving this 
problem; see in particular Section 4 of that paper. 

One of the keys to sparse regularization is the choice of the dictionary and proper tuning of the 
sparsity constraint. Glimpse2D adopts a wavelet based dictionary for $\mathbf{\Phi}$, combining a 
multi-scale starlet dictionary \citep{SFM.2007} along with a Battle-Lemari\'e wavelet to help constrain 
the smallest scales. We find this combination of dictionaries very well suited to the convergence signal 
at the cluster scale, and this prior can lead to near-perfect reconstruction in the absence of noise. To 
tune the sparsity constraint based on the level of noise on different scales and at different locations 
in the field, Glimpse2D adopts a weighted $\ell_1$ norm, through the weight factor $\bm{w}$ in
Eq. \eqref{eq:algo_rec_non_lin_linearised_z}. This factor is computed empirically by propagating
randomized galaxy ellipticities through to wavelet coefficients and estimating the resulting scale 
and position dependent noise standard deviation. Using this scheme, the sparsity constraint in Glimpse2D 
is tuned by a single parameter $\lambda$. 

The regularization parameter $\lambda$ controls the trade-off between fitting the observed data and
enforcing sparse solutions. A large value will provide a solution containing features that can be
considered as real with a very high probability, but faint features may be lost. A small value preserves
the smallest features, but some of them may be due to the noise.
There are empirical motivations, however, which guide our choices in this work
and that we describe further in Section \ref{sec:results}. 
See also the Appendix, where we present numerical experiments testing the Glimpse2D algorithm
on the simple case of a known halo mass profile in the context of A520-like noise. We find that a typical 
value of $\lambda$ to obtain a good mass reconstruction is $\sim$3 for noisy data.

Compared to other approaches, Glimpse2D is better able to preserve small-scale information by 
avoiding the need to bin the shear measurements before solving for convergence.
Another benefit is that the algorithm is able to incorporate flexion measurements of the 
individual galaxies into the reconstruction when available, which significantly improves the 
recoverability of small-scale features. As there are no flexion measurements for A520, we do not 
use this feature in the present work. However, even without flexion, Glimpse2D is still an 
extremely powerful reconstruction technique due to its sparsity-based regularization scheme. A 
good illustration of this is Figure 6 in \citet{LSL.etal.2016}, where sparse regularization
yields a near-perfect reconstruction on a noiseless inversion problem, despite 93\% missing
pixels in the input shear field.

\section{Results}\label{sec:results}
In this section we present the Glimpse2D mass map reconstructions of A520 from both the 
weak-lensing catalogs of \cta{C12} and \cta{J14}. To make meaningful comparisons
between our results and the previously published maps, we perform the reconstructions for each data
set assuming the cosmology used in its respective paper. Both papers assumed a flat $\Lambda$CDM 
cosmological model with a present-day Hubble constant of $H_0=70$ km s${}^{-1}$ Mpc${}^{-1}$.
As for the total matter density, \cta{C12} used $\Omega_\mathrm{m}=0.27$, while
\cta{J14} used $\Omega_\mathrm{m}=0.3$. This difference is unimportant for our purposes,
as it only slightly scales the amplitude of the lensing signal between the two analyses. 

For all of the results that follow, we run Glimpse2D using the same configuration parameters. 
The resolution of the reconstructed mass maps can be as high as we wish, since we are not limited 
by a prior binning of the data. The cost of smaller pixels, however, is increased computation 
time. We set the pixel size to 0.033 arcmin as a balance between resolution and speed. 

We choose to use 8 wavelet scales, since this is the maximum number allowed by the size
of our output images. In the multi-scale regularization step of the algorithm, the wavelet function 
is dilated by a factor of 2 at each new resolution level. For our $256\times256$ pixel maps, this
corresponds to a maximum of $\log_2 256=8$ possible scales. The noise of the coarsest resolution scale
can affect the solution in principle, since it does not have zero mean and is therefore not 
thresholded like the smaller scales. On large enough scales, however, the noise is typically 
negligible, so we should achieve the best solution by using the largest number of wavelet scales 
permitted by the image size. Using a smaller number could negatively impact the results by
allowing the potentially higher noise level on smaller scales to enter the reconstructions.

As Glimpse2D is an iterative solver, we must also specify the 
number of iterations. We set this to 500, since we have verified that the algorithm
converges by this point for the range of $\lambda$ values of interest for both data sets.
Following \citet{LSL.etal.2016}, we use 5 re-weightings (see Section 3.2) in order to 
help correct for the possible bias induced by the $\ell_1$ sparsity constraint.
Finally, the code is run with a positivity constraint on the reconstructed $\kappa$ in order to 
focus only on the overdense peaks. While this effectively adds a mass sheet to the reconstructions,
since we do not compute (aperture) masses of the substructures, the result we present are not 
affected by this choice.

\begin{figure*}
\centering
\includegraphics[width=18cm]{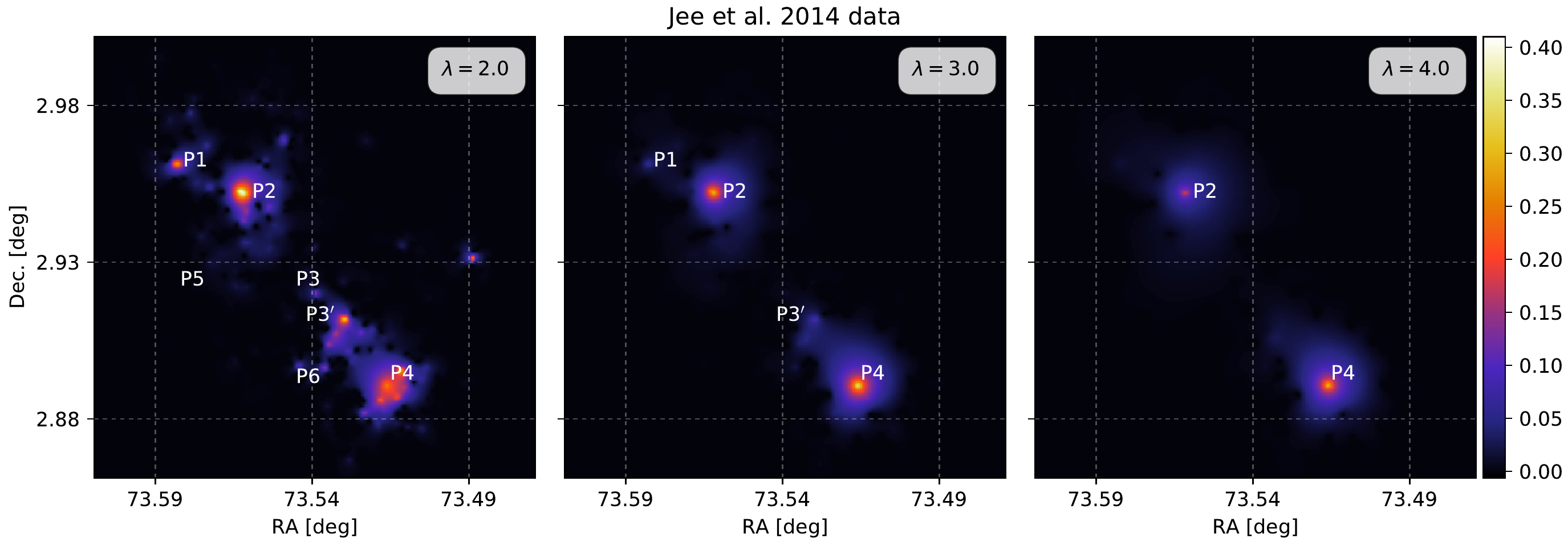}
\caption{Surface mass reconstructions from \cta{J14} data for regularization parameters $\lambda=2.0$, 
         3.0, and 4.0. Labels P1--P6 are the same as in Figure \ref{fig:fig1}. The mass maps agree well with
         the \cta{C12} data overall, but there are important differences. In particular, the P5 structure
         seen in \cta{C12} is missing here, while the dark core at P$3'$ more prominent.}
\label{fig:fig2}
\end{figure*}

\subsection{C12: Magellan + HST/ACS}\label{ssec:recs}
Figure \ref{fig:fig1} shows three reconstructions of the \cta{C12} data corresponding to 
regularization parameters $\lambda=2.0$, 3.0, and 4.0. Recall that a higher $\lambda$ value 
promotes a sparser solution and thus more effectively suppresses the noise and lower significance 
features. From left to right, the maps visually exhibit the expected trend in terms of the number 
and amplitudes of mass peaks. We label the locations of structures identified in \cta{C12} and 
\cta{J14} by the same numbering scheme used in those studies. It is clear that structures P1, P2, 
and P4 are the most prominent, as they appear in all three reconstructions, while the other peaks 
that show up at $\lambda=2.0$ have disappeared at $\lambda=3.0$ and above. P4 has the highest 
amplitude, followed by P2, which is consistent with the column mass estimates given in \cta{C12} 
(see their Tables 1 and 2). 

Of particular interest are the locations P3 and P$3'$, the previous and updated detection 
positions, respectively, of the dark peak. In agreement with \cta{C12}, the P3 region shows no 
indication of the dark core reported by \cta{M07} and \cta{J12}. On the other hand, there is 
indeed a visible peak near P$3'$ in the $\lambda=2.0$ map. It has a smaller size and lower 
amplitude compared to the other structures, apart from P6. It is also not immediately clear that 
this constitutes a true mass concentration rather than simply a noise fluctuation. 
A similarly-sized peak appears just south of P$3'$ in the direction of P6, as do two others in the
outskirts of the cluster. Among these, however, only P$3'$ is still visible up to $\lambda=2.3$.
The others disappear between $\lambda=2.1$ and 2.2, suggesting that they are likely spurious noise 
peaks, whereas P$3'$ is a potentially real, albeit low-amplitude feature. We return to the issue 
of quantifying the significance of the P$3'$ detection in Section \ref{ssec:significance}.

Another interesting aspect of the reconstructions is in the morphologies of P2 and P4. Each region
shows multiple connected peaks at $\lambda=2.0$ that essentially converge into a single peak by 
$\lambda=4.0$. Some indication of substructure is still apparent at P4 for $\lambda=4.0$, however. 
The fact that the P2 centroid remains stable as a single peak between $\lambda=3.0$ and 
$\lambda=4.0$ strongly suggests that the secondary peak seen at $\lambda=2.0$ is caused by noise, 
even though it appears at higher relative amplitude in that map. On the other hand, the presence 
of the pair of peaks at P4 even in the higher $\lambda$ maps could reflect the ability of 
Glimpse2D to recover small-scale information afforded by the high resolution of the reconstructions.

\begin{deluxetable}{l c c} 
\tabletypesize{\footnotesize} 
\tablecolumns{3} 
\tablewidth{0.45\textwidth} 
\tablecaption{Maximum values of regularization parameter $\lambda$ for which the substructures 
              are detectable in the two data sets. \label{tbl:max_lambda}} 
\tablehead{\colhead{Substructure} & \colhead{C12 $\lambda_\mathrm{max}$} & \colhead{J14 $\lambda_\mathrm{max}$}}
\startdata
\phm{\quad\qquad}P1    &  $4.5$   & 3.0 \\[0.06cm]
\phm{\quad\qquad}P2    &  $>5.0$  & $>5.0$ \\[0.06cm]
\phm{\quad\qquad}P3    &  ND      & 2.5 \\[0.06cm]
\phm{\quad\qquad}P3$'$ &  2.3     & 3.6 \\[0.06cm]
\phm{\quad\qquad}P4    &  $>5.0$  & $>5.0$ \\[0.06cm]
\phm{\quad\qquad}P5    &  2.3     & ND \\[0.06cm]
\phm{\quad\qquad}P6    &  2.0     & 3.0
\enddata
\tablecomments{ND stands for no detection.} 
\end{deluxetable}

\subsection{J14: CFHT + HST/ACS}
In Figure \ref{fig:fig2} we show three reconstructions of the \cta{J14} data for the same $\lambda$
values as in Figure \ref{fig:fig1}, with the relevant substructures again labeled P1--P6. The maps
agree well overall with the entropy-regularized maximum likelihood reconstruction presented in
\cta{J14} (see their Figure 5). For $\lambda=2.0$ it is easiest to see the correspondence between
the shapes and relative positions of the Glimpse2D substructures and those of \cta{J14}. In terms 
of size and amplitude, the most prominent features in both are P2 and P4, followed by P1 and 
P$3'$. 

P5 is curiously missing from all of our reconstructions, whereas \cta{J14} found it to 
have a projected mass larger than that of P1. Given differences in the sharpness of the two peaks, 
seen both in our $\lambda=2.0$ map of Figure \ref{fig:fig1} as well as the mass map of \cta{J14}, 
we have explored the possibility that our choice of the number of wavelet scales might be
influencing the result.
We have verified that decreasing the number of scales from 8 to 5, where now the size of the 
broadest wavelet function corresponds to an angular scale of $\sim$1 arcmin, does not produce a peak at P5.
If indeed there is a significant mass peak in the data at this location, our algorithm appears 
to be insensitive to it. This would indeed be surprising, however, as we have not seen such 
behavior in tests on simulations.

With the \cta{J14} data, the dark core at P$3'$ appears as a distinct peak comparable in size to
P1. It is still detectable at $\lambda=3.0$, unlike for the \cta{C12} data, where instead P1 remains
as the third prominent substructure at that level. Some possible fragmentation of P2 and P4 is 
noticeable at $\lambda=2.0$ in the \cta{J14} data, although it is less apparent than in \cta{C12}.
Finally, there does appear to be a peak at the location of P3 at $\lambda=2.0$, but it is not 
distinguishable from the many other low-amplitude noise peaks scattered across the field at that
level, which all vanish for $\lambda\approx 2.5$ and above. We therefore confirm the conclusion of \cta{C12}
and \cta{J14} that there is no significant dark mass concentration at the P3 location.

We summarize the detectability of the various labeled peaks in Table \ref{tbl:max_lambda} for the two
data sets. Reconstructions were made for $\lambda\in[2.0, 5.0]$ in increments of $\Delta\lambda=0.1$.
Substructures not detected within this range are marked as ND.

\begin{figure*}
\centering
\includegraphics[width=18cm]{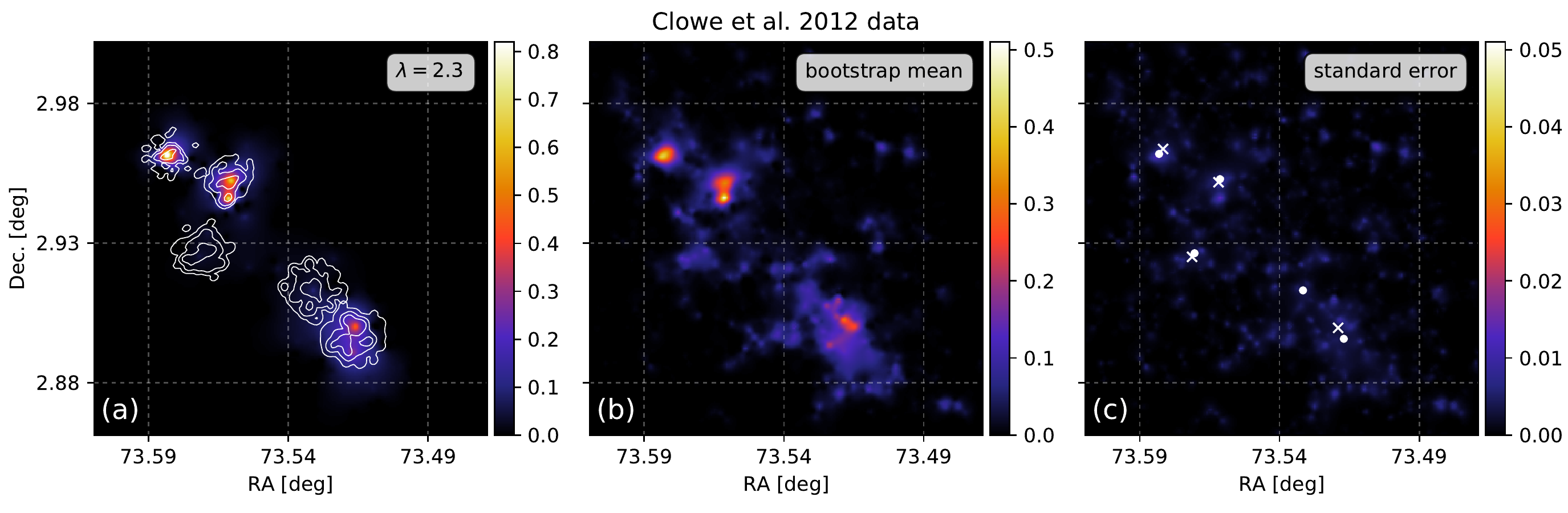}
\caption{Bootstrap analysis of \cta{C12} data with $\lambda=2.3$. (a) Position uncertainties are computed
         from $N=1000$ bootstrap reconstructions and are indicated by contour lines surrounding each 
         centroid. Outward from the centers, the contours enclose 68\%, 95\%, and 99.7\% of the centroid
         detections. The background image is the $\lambda=2.3$ Glimpse2D reconstruction of the original
         catalog for reference. (b) Mean map of the bootstrap reconstructions. Many more features are
         present here compared to in (a), both due to noise peaks as well as shifting of the primary
         centroid positions. (c) Pixel-wise standard error on the bootstrap mean. The highest
         error regions track the features seen in panel (b), although the overall amplitude is low.
         Locations of the major substructures determined as maxima of the mean bootstrap map are marked
         as filled circles. Centroids reported in \cta{C12} are shown with an $\times$ for comparison.}
\label{fig:C12bootstraps}
\end{figure*}

\subsection{Peak position uncertainties}\label{ssec:bootstrap}
To study the uncertainties of our mass reconstructions, we perform a bootstrap analysis by 
generating $N=1000$ re-sampled (with replacement) catalogs from both data sets and running 
Glimpse2D on each to obtain $N$ mass maps.
We can study the positional uncertainties of the primary substructures for a given $\lambda$ 
by examining the distributions of their detected locations in a large number of bootstrapped
reconstructions. 

We compute these uncertainties by considering a circle of radius 150 kpc centered on each structure
and taking the closest peak to the centroid in each bootstrap detected within the interior. 
This circular search area coincides with the aperture size used in both \cta{C12} and \cta{J14} 
and also corresponds approximately to the typical positional variability of a P4-like structure 
in the noise simulations studied in the Appendix.
We then determine three contour levels surrounding each mass peak that enclose 68\%, 95\%, and 
99.7\% of the detections from the $N$ resampled maps. For both data sets, we have verified that 
the contours do not change whether we take the reference centroid positions (centers of the 
circles) to be the peaks from the original catalog reconstruction or as the local maxima of the 
bootstrap mean. 

To fix the $\lambda$ value for each data set, we seek a level high enough that the
frequency of false detections due to noise does not strongly impact the resulting uncertainty
contours. 
This will occur if many noise peaks appearing within the centroid search areas are mistaken for
the true centroid. We also want to be able to detect the primary substructures of interest in
each boostrap map, meaning $\lambda$ should not be so high as to suppress reconstruction of 
the lower--amplitude centroids. We therefore set $\lambda$ to be the highest value in which 
P1, P2, P3$'$, and P4 are all detectable based on the original data reconstructions
(cf. Table \ref{tbl:max_lambda}). This is $\lambda=2.3$ for the \cta{C12} data and $\lambda=3.0$ 
for the \cta{J14} data.

\begin{deluxetable}{l c c c c} 
\tabletypesize{\footnotesize} 
\tablecolumns{5} 
\tablewidth{0.47\textwidth} 
\tablecaption{Centroid properties from bootstrap analysis of \cta{C12} data.\label{tbl:C12}} 
\tablehead{             & R.A. & Dec. & $\Delta$R.A. & $\Delta$Dec. \\[-0.1cm]
           Substructure &  &  &  & \\[-0.1cm]
                        & (h : m : s) & ($^\circ$ : $'$ : $''$) & ($''$) & ($''$)}
\startdata
\phm{\qquad}P1    & 04:54:19.94 & $+$02:57:42.66 & $+5.16$ & $-6.43$ \\[0.06cm]
\phm{\qquad}P2    & 04:54:14.71 & $+$02:57:10.19 & $-1.92$ & $+3.94$ \\[0.06cm]
\phm{\qquad}P3$'$ & 04:54:07.58 & $+$02:54:46.87 & -- & -- \\[0.06cm]
\phm{\qquad}P4    & 04:54:04.08 & $+$02:53:44.48 & $-7.35$ & $-14.1$ \\[0.06cm]
\phm{\qquad}P5    & 04:54:16.90 & $+$02:55:34.75 & $-3.21$ & $+4.66$
\enddata
\tablecomments{Differences $\Delta$ are computed by subtracting the published \cta{C12} values 
               from the local maxima of our mean bootstrap map. \cta{C12} did not study the 
               P3$'$ location, so we give only our determined centroid position.} 
\end{deluxetable}

\begin{figure*}
\centering
\includegraphics[width=18cm]{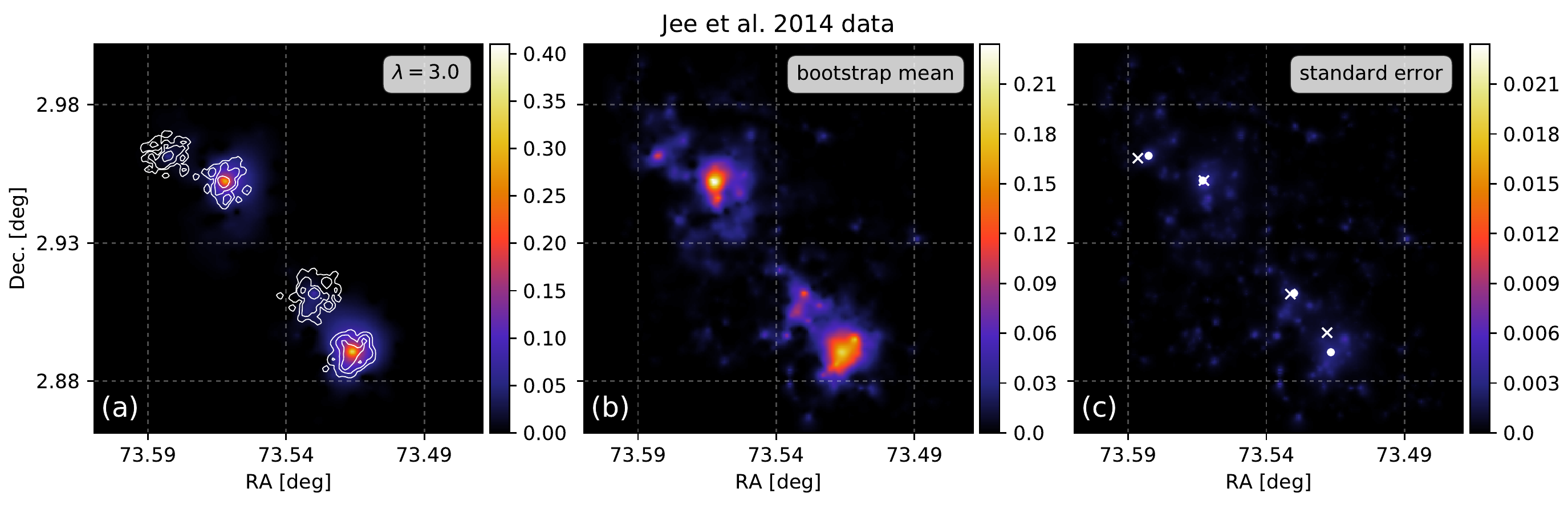}
\caption{Bootstrap analysis of \cta{J14} data with $\lambda=3.0$. Panels are analogous to those
         in Figure \ref{fig:C12bootstraps}. }
\label{fig:J14bootstraps}
\end{figure*}

We show results from the \cta{C12} bootstraps in Figure \ref{fig:C12bootstraps}. In panel (a) we plot 
the centroid uncertainty contours for substructures P1, P2, P3$'$, P4, and P5, which are overlaid on 
the original catalog reconstruction for reference. The tighter contours of the P1 and P2 
substructures indicate that their positions are the most stable, whereas P3$'$ and P4 show more 
variability. This is not too surprising considering the relative strengths of these peaks as seen 
in the $\lambda=2.0$ map of Figure \ref{fig:fig1}. The uncertainty areas we derive are comparable 
to, but slightly smaller than those obtained by \cta{C12} in a similar bootstrap analysis (see 
their Figure 4). The most notable aspect of the contours is that the positional variations 
of P3$'$ and P4 are large enough that their 3$\sigma$ regions almost overlap. Although the 1 and 
2$\sigma$ regions remain clearly distinct between the centroids, their outer contours are 
approaching the 150 kpc limit and thus approaching each other. We contrast this to the \cta{J14}
results in Figure \ref{fig:J14bootstraps}.

Shown in panel (b) of Figure \ref{fig:C12bootstraps} is the mean of the bootstrap reconstructions. 
Comparing to the original reconstruction in (a) and to those of Figure \ref{fig:fig1}, the 
bootstrap mean exhibits significantly more structure overall due to shifting of the primary
centroid positions as well as the appearance of noise peaks. The variability of the centroid 
locations is reflected in the contours in panel (a) and depends on the strength of the lensing
signal. The standard error on the mean (pixel-wise) is shown in panel (c). The variability
is around 10\% the amplitude of the mean, and a clear correlation with panel (b) can be seen.
We attribute the numerous additional noise features appearing in (b) primarily to the low value 
of regularization parameter used. 

Finally, we compare our measured centroid positions (marked by filled circles) with those reported in 
\cta{C12} (marked by $\times$'s) in panel (c) as well. There is no $\times$ at P3$'$,
since the updated dark core position was not available to \cta{C12} when they performed their
analysis. Our centroid locations are in excellent agreement with those of \cta{C12}. Right
ascension (R.A.) and declination (Dec.) values for each are given in Table \ref{tbl:C12}, along
with the differences between our coordinates and theirs. The largest discrepancy, which occurs
for P4, is an angular separation of smaller than 16$''$.

For the \cta{J14} data, we carry out an analogous bootstrap analysis to the one we did for \cta{C12} described
above. Results are presented in Figure \ref{fig:J14bootstraps}. Panel (a) shows the uncertainty contours
computed from the $N=1000$ bootstraps overlaid on the original $\lambda=3.0$ reconstruction. In this
case we focus only on P1--P4, since we do not detect P5 as a prominent feature. The contours are well 
localized on their respective centroids and have overall smaller sizes to those in Figure 
\ref{fig:C12bootstraps}. In contrast to our results with \cta{C12} data, however, the P3$'$ is
clearly seen as a substructure distinct from P4. This agrees with expectations based on the
reconstructions in Figure \ref{fig:fig2}.

In panel (b), we see again that the bootstrap mean exhibits more features than its counterpart in 
(a), although it is not as noisy compared to panel (b) of Figure \ref{fig:C12bootstraps}. The primary
reason for this is the higher $\lambda$ value used. Another contributing factor is also likely 
the higher source number density of the \cta{J14} catalog. It is interesting to notice the multiple
peaks visible within the P4 region. These coincide with the hints of additional low-amplitude peaks
seen in the $\lambda=2.0$ map of Figure \ref{fig:fig2} but not for $\lambda=3.0$. 

As was the case in Figure \ref{fig:C12bootstraps}, the overall error structure shown in panel (c) 
correlates well with the bootstrap mean. We mark the centroid positions from our bootstraps as 
filled circles and compare to the reported locations in \cta{J14}, which are marked by an $\times$. 
Our centroids are again in good agreement with the published locations, in particular P1, P2, and 
P3$'$, which are all within an angular separation of 14$''$. The largest discrepancy is now P4 
with a separation of $\sim26''$, similar to that of P4 in the \cta{C12} analysis. Centroid
coordinates and differences with the \cta{J14} values are given in Table \ref{tbl:J14}.

An important difference between our analysis and that of \cta{J14} is that we do not use the 
bootstraps to determine the significance of the detected structures.
This can usually be done, 
for example, by considering the number of times a particular peak is detected above a given 
threshold out of the $N$ re-sampled maps. One problem with this approach for Glimpse2D lies 
in setting the detection threshold in a meaningful way. Using a global $k\sigma$, $k=1,2,3...$
level, where $\sigma$ is the standard deviation of the bootstrap map does not work, since the interpretation
of this $\sigma$ is not clear. Furthermore, our maps are generated for a chosen value of $\lambda$, which 
essentially imposes a prior threshold on the result, below which structures are not reconstructed and which 
varies depending on the properties of the particular realization.
Instead then, we have set the threshold for detection low enough that a peak within the circle of 
radius 150 kpc is found in all of the bootstrap maps.
We address the question of significance in the following section.

\begin{deluxetable}{l c c c c} 
\tabletypesize{\footnotesize} 
\tablecolumns{5} 
\tablewidth{0.47\textwidth} 
\tablecaption{Centroid properties from bootstrap analysis of \cta{J14} data.\label{tbl:J14}} 
\tablehead{             & R.A. & Dec. & $\Delta$R.A. & $\Delta$Dec. \\[-0.1cm]
           Substructure &  &  &  & \\[-0.1cm]
                        & (h : m : s) & ($^\circ$ : $'$ : $''$) & ($''$) & ($''$)}
\startdata
\phm{\qquad}P1    & 04:54:19.82 & $+$02:57:41.54 & $-14.1$ & $+3.14$ \\[0.06cm]
\phm{\qquad}P2    & 04:54:15.13 & $+$02:57:09.26 & $+1.58$ & $+0.06$ \\[0.06cm]
\phm{\qquad}P3$'$ & 04:54:07.19 & $+$02:54:42.42 & $-4.82$ & $+1.12$ \\[0.06cm]
\phm{\qquad}P4    & 04:54:03.99 & $+$02:53:25.53 & $-4.90$ & $-25.5$
\enddata
\tablecomments{Differences $\Delta$ are computed by subtracting the published \cta{J14} values from
               the local maxima of our mean bootstrap map.} 
\end{deluxetable}

\subsection{Significance of the dark core}\label{ssec:significance}
The results of the previous sections indicate that there is indeed a peak visible at P3$'$ in the 
\textit{HST}/ACS A520 data, at least for some levels of the regularization parameter.
We now address the question of quantifying the significance of this detection.

As discussed above, the regularization parameter functions as a detection threshold in such a way 
that at a given $\lambda$, only features above a particular amplitude relative to the local 
noise level are reconstructed. The higher the $\lambda$ value, the more effectively the algorithm 
removes low significance features. We can therefore think of $\lambda$ as providing a natural
significance scale in the following sense. We interpret features appearing in a $\lambda=\lambda_1$ 
map but not in a $\lambda=\lambda_2$ map, where $\lambda_1 < \lambda_2$, as being less significant than
features appearing in both maps. We can say, then, that P2 and P4 are the most significant cluster 
substructures in both data sets we have studied, in agreement with the \cta{C12} and 
\cta{J14} analyses. Furthermore, the peak at P1 appears at higher significance in the \cta{C12} data
than in the \cta{J14} data, whereas the reverse is true for P3$'$.

To make this intuition more quantitative, we aim to associate the presence of a peak in a particular 
$\lambda$ map to a $k\sigma$ detection level based on noise simulations of the data. 
We first generate $N$ Monte Carlo simulations of the
original data sets in which both the positions and orientations of the galaxies are randomized. 
Each resulting noise simulation occupies the same footprint on the sky as the catalog it derives
from (see Figure \ref{fig:galaxydensity}) and contains the same number of galaxies. For the 
\cta{C12} data, the Magellan and ACS galaxies are initially treated separately and then combined 
into a single catalog in order to respect the original galaxy number density.

\begin{figure}
\centering
\includegraphics[width=8.5cm]{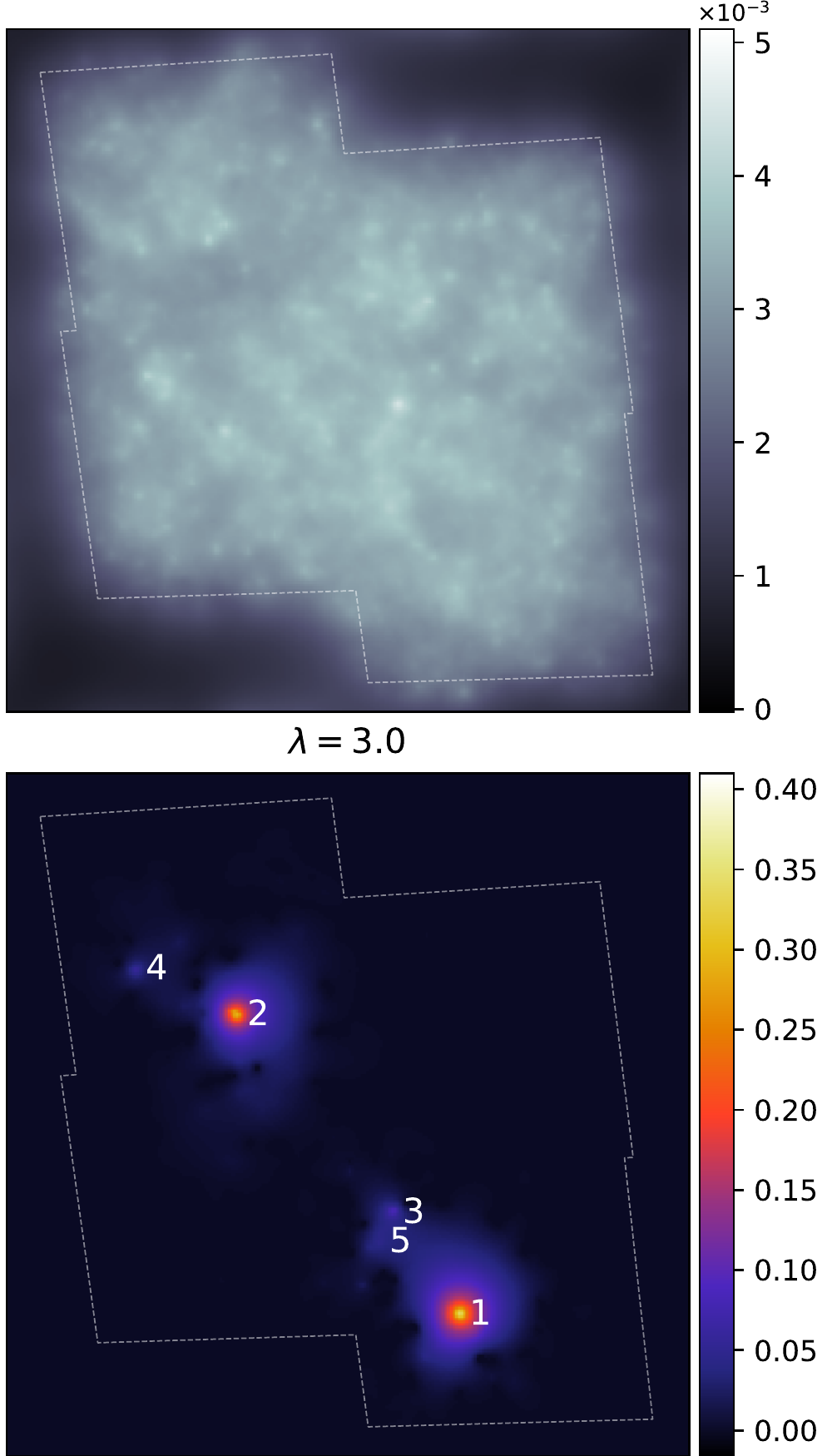}
\caption{Example results from $N=1000$ noise simulations of the \cta{J14} data with $\lambda=3.0$.
         The upper panel shows the mean of the noise maps, which exhibits a high-amplitude central
         region relative to the border, coinciding with the ACS mosaic field of view. The mass reconstruction
         of the original catalog is shown in the lower panel, with filled circles marking the positions of 
         the highest peaks detected. The accompanying numbers indicate rank ordering of the peaks by amplitude.
         Peaks detected in the noise simulations are used establish the significance of these features, as
         explained in the text.}
\label{fig:J14noise}
\end{figure}

We run Glimpse2D on $N=1000$ such noise simulations and consider the number and amplitudes of peaks 
detected in each reconstructed map. As an example visualization, we show in the upper panel of 
Figure \ref{fig:J14noise} the mean of the noise maps for \cta{J14} data with $\lambda=3.0$. 
The polygonal pattern filling the central region reflects the shape of the ACS field. The fact that this region is
higher amplitude with respect to its surroundings indicates that the majority of the noise peaks reconstructed 
in each simulation fall within the boundary of the source galaxy field. Compared with the data reconstruction
(lower panel), the overall noise level is about two orders of magnitude lower than the amplitudes of the P2 
and P4 mass peaks. Since we have randomized galaxy positions as well as their orientations, spatial 
fluctuations seen in the mean noise map are merely statistical.

The mass reconstruction of the original catalog is shown in the lower panel of Figure \ref{fig:J14noise}, 
again for $\lambda=3.0$. The highest peaks detected in the map are marked with 
numbers indicating rank ordering by amplitude, where 1 is the highest.
After numbers 1 and 2, which correspond to substructures P4 and P2 (cf. Figure \ref{fig:fig2}), the 
remaining peaks are difficult to discern by eye given their low relative amplitudes. 
The third highest peak, corresponding to the dark core location at P3$'$, is about 
24\% the amplitude of the highest.

An interesting feature of the P3$'$ region is that two peaks in fact appear at comparable 
amplitude, numbers 3 and 5. This is apparent as well in the position uncertainties in 
Figure \ref{fig:J14bootstraps}, where the contours centered on the third peak extend to 
enclose the fifth highest peak toward the southeast. The two peaks remain at similar relative 
amplitudes up to $\lambda=3.5$, near the level where they vanish from the reconstructions.
This could be a hint that the peaks are in fact connected, i.e., that the region actually 
contains an elongated mass structure. If this is the case, it is possible that a different 
wavelet dictionary from the isotropic starlet we have used in Glimpse2D might be better at 
revealing its morphology. Finally, we note that unlike P2 and P4, whose rank ordering 
remains the same for all $\lambda$ values we have considered, the secondary peak associated with 
P3$'$ (number 5) switches places with P1 (number 4) in the ordering for $\lambda=3.5$, 
meaning that its detectability actually increases slightly relative to the primary P3$'$ peak 
with increasing $\lambda$.

For this $\lambda$, we can assign a significance level to the peaks by determining the probability $p$ of
a pure noise peak to occur at or above the true peak amplitude in the data reconstruction. We can then
convert $p$ into a $k\sigma$ value by finding the $k$ satisfying $1-p=\erf(k/\sqrt{2})$, as the right-hand side
represents the fraction of a Gaussian distribution lying within $k\sigma$ of the mean. We focus on the 
primary subclusters P2 and P4, since they constitute the strongest isolated detections in the field that also
remain consistent across the various $\lambda$ values.
Considering first the highest peak P4, the number of noise maps in which a peak appears at or above the P4 
height is 37, corresponding to a significance of $2.1\sigma$. For P2, we find 50 such false detections,
corresponding to $2.0\sigma$. Given the low amplitudes and inconsistent ordering of the remaining peaks, 
the interpretation of their associated false detection rates is not as straightforward. 
We can conclude, however, that their significance must be less than that
of P2, or $2.0\sigma$.

We can update the detection significance values by repeating the analysis for larger $\lambda$.
For structures with amplitudes that are still high enough to be detected (i.e., P2 and P4) the values 
represent a new lower bound on their significance. For example, with $\lambda=3.5$, P4 and P2 appear now at
$3.0\sigma$ and $2.5\sigma$, respectively. At $\lambda=5.0$, no noise peaks appear above either the P4 or P2 
amplitudes out of 1000 realizations, implying that their detection significances are at least $3.3\sigma$. We
expect this trend to continue at still larger $\lambda$ values, but the computation would require more than
the number of noise realizations we have at hand.

Carrying out the analogous process on the \cta{C12} data with $\lambda=3.0$, we find that the most significant
substructures are P1, P2, and P4, with corresponding significance levels of 1.6$\sigma$, 1.5$\sigma$, and
1.4$\sigma$ for this $\lambda$.
This could be anticipated from the data reconstructions shown in Figure \ref{fig:fig1} and from Table
\ref{tbl:max_lambda}. While these limits would certainly increase by analyzing simulations at higher $\lambda$,
it is clear already at this level that the significance of the other peaks must be less than the minimum of 
the highest three, or 1.4$\sigma$. This includes the dark peak region, which does not appear 
for $\lambda=3.0$. 

The ordering of peak amplitudes in the \cta{C12} data represents an interesting departure
from the \cta{J14} data. In particular, the significance of P1 is elevated relative to P2 and P4 at this
$\lambda$, but this would not remain the case at $\lambda=4.0$ and above based on the Figure \ref{fig:fig1}
reconstructions. The prevalence of P1 here compared to in the \cta{J14} data could be due to a dilution of the 
lensing signal at this location by the numerous faint galaxies included in the \cta{J14} catalog but not in 
the \cta{C12} catalog.

Finally, it is important to point out some relevant considerations in our determination of peak 
significance levels from pure noise simulations. The first is that in our calculation, we have assumed that all
locations in the reconstructed field are equally probable. For high-amplitude peaks that are clearly detected, 
such as P2 and P4, we expect this assumption to hold well. However, there is likely some degree of 
dependence on the probability of reconstructing P3$'$ due to the presence of P4, given their proximity.
Another caveat is that we have performed only a weak-lensing analysis in this work. We have not considered, 
for example, peaks in the X-ray gas map that are known to concide with the P3$'$ location \citep{WMG.2016}. 
A joint study could be useful here to determine the probability of obtaining a peak simultaneously in both 
X-ray and weak-lensing maps.

\section{Summary and Conclusion}\label{sec:conclusion}
We have analyzed the A520 merging cluster using a novel sparsity-based mass-mapping code called Glimpse2D. 
The goal of the re-analysis was twofold. The first was to test the Glimpse2D algorithm on real data, since it
has been shown to perform well on simulations of \textit{Hubble} ACS-like data, but has not been applied
to real observations until now.
Second, given the reported detection of a significant but anomalously high mass-to-light ratio structure 
in the A520 system, we sought to carry out a separate study using an independent mass inversion 
technique to assess the presence of the peculiar dark core. 

We obtained the source galaxy catalogs with shape measurements from two different groups who have 
studied the cluster. The first catalog is that of \cta{C12}, which consists of galaxies derived from
a combination of Magellan images and \textit{HST}/ACS mosaics. The second catalog is from \cta{J14},
which derives from the same \textit{HST}/ACS images that were used in \cta{C12}. The resulting catalogs
are different, namely in the number density of galaxies in their common regions of sky coverage
due to different reduction pipelines. The \cta{J14} catalog contains nearly twice the number of galaxies
as that of \cta{C12}.

Our mass reconstructions obtained from running Glimpse2D on both data sets were in good agreement overall 
with those presented in their respective papers. The two main subcluster components were reconstructed along 
the same merger axis inferred by the previous studies. In particular, for the \cta{C12} reconstructions, we 
found the substructures labeled P1, P2, and P4 to be the most prominent, as they were detectable up 
to the highest value of regularization parameter ($\lambda=4.0$) used. 
A peak was visible at the P3$'$ position in the 
$\lambda=2.0$ reconstruction, but not above $\lambda=2.3$. On the contrary, for the \cta{J14} 
reconstructions, P3$'$ was undetectable only in the maps with $\lambda > 3.5$, indicating its higher likelihood 
of being a real structure. The P1 peak, however, appeared much less prominent than P2 and P4, and somewhat
less than P3$'$ as well.

To study the positional uncertainties on the mass centroids, we performed bootstrap analyses on
both data sets. We examined the variability in the detected locations of P1--P5 for the \cta{C12} data 
and P1--P4 for the \cta{J14} data, finding that the peaks were well localized in nearly all cases. The
68\%, 95\%, and 99.7\% confidence contours around each centroid remained separate from the others, 
except those for P3$'$ and P4 in the \cta{C12} data, which exhibited overlap of their outer regions. 
It is thus reasonable to interpret the P3$'$ peak as a distinct structure (whether as a spurious 
noise peak or indeed as a real concentration of mass) in the \cta{J14} data, but not in the \cta{C12} data.

We established the significance of the peak detections by generating Monte Carlo noise
simulations of the two catalogs at various $\lambda$ values. Assuming independence of the peaks, i.e., that all
locations in the field are equally likely, we placed an upper limit on the significance of the P3$'$ dark 
peak at $2.0\sigma$ using the \cta{J14} data and $1.4\sigma$ using the \cta{C12} data.
This is substantially less than the greater than $6\sigma$ rejection of the constant mass-to-light ratio hypothesis 
that was reported by \cta{J14}. We showed then that in neither case can we confirm the detection of a dark 
peak, the reality of which would be problematic within the current understanding of dark matter as an
effectively collisonless particle.

We finally note that we have not carried out a full mass-to-light ratio analysis of the 
A520 data, since we are currently unable to do this independently with Glimpse2D. The issue of
measuring a peak mass in excess of what is expected---based on, for example, a hypothesis of constant
mass to light---is related to, though logically distinct from assessing the significance of a 
peak appearing at that location. However, \cta{J14} reported a projected mass for P3$'$ consistent
with that of P2 and only about 7\% smaller than that of P4. Given that we have detected both P2 
and P4, but not P3$'$, with Glimpse2D at high confidence, we cannot confirm by our analysis that 
there is indeed such an anomalous peak at the purported location.

In the spirit of presenting reproducible research, we have made the galaxy catalogs used in this work to
produce our mass reconstructions available on the CosmoStat
website at \url{http://www.cosmostat.org/software/glimpse/}. 
Configuration files, output maps, and instructions for running Glimpse2D on the catalogs have been included
as well.

\acknowledgments
We thank Douglas Clowe, Richard Massey, James Jee, and Henk Hoekstra for their insightful 
comments on the draft and for kindly agreeing to share their weak-lensing catalogs with us.
We would also like to thank Florent Sureau, Samuel Farrens, and Ming Jiang for useful 
discussions during the preparation of the paper. 
This work is supported in part by \textit{Enhanced Eurotalents}, a Marie Sk{\l}odowska-Curie 
Actions Programme co-funded by the European Commission and Commissariat {\`a} l'{\'e}nergie 
atomique et aux {\'e}nergies alternatives (CEA).
The authors acknowledge the Euclid Collaboration, the European Space Agency, and the support of 
the Centre National d'Etudes Spatiales (CNES).
This work is also funded by the DEDALE project, contract no. 665044, within the H2020 Framework 
Program of the European Commission.

\appendix
We present here the results of numerical experiments carried out on the A520 data to test 
Glimpse2D in the current setting. As discussed in Section \ref{ssec:glimpse}, the regularization
parameter $\lambda$ controls the sparsity constraint on the mass map reconstructed by Glimpse2D.
Since we need to fix the trade-off between smoothness and sensitivity through the choice of
$\lambda$, we aim to guide our intuition empirically by testing the algorithm on simulations 
representative of the data. Algorithm configuration parameters here are the same as 
those used to obtain the main text results.

\begin{figure*}
\centering
\includegraphics[width=17.8cm]{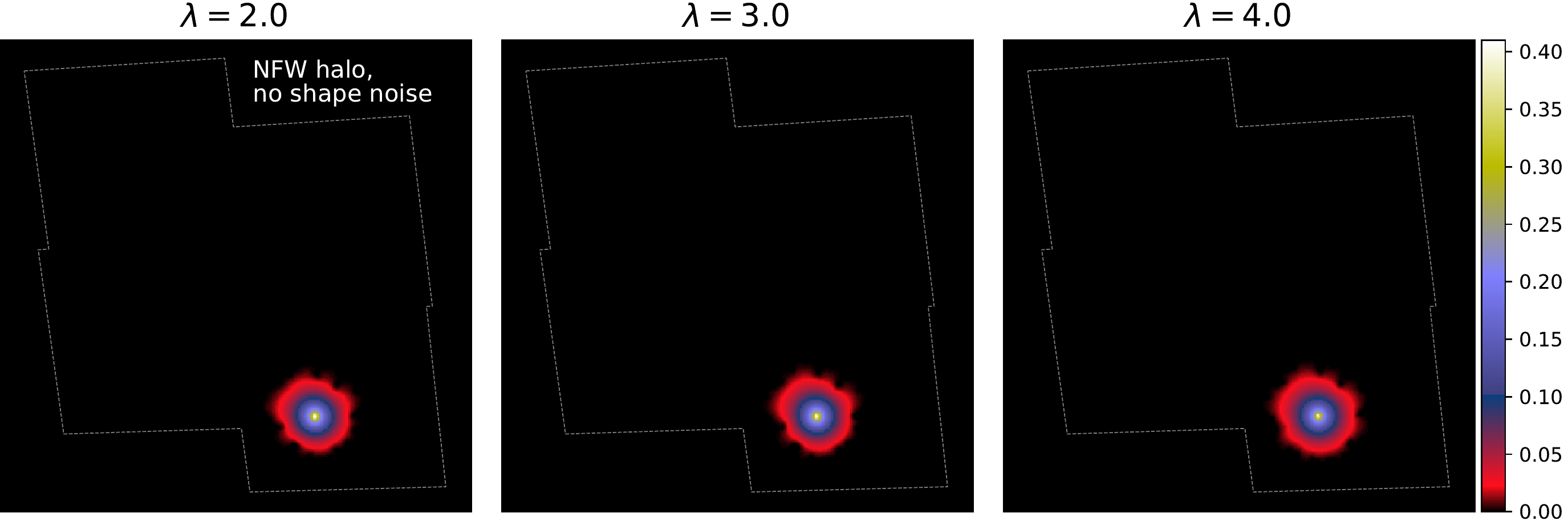}\\
\includegraphics[width=17.8cm]{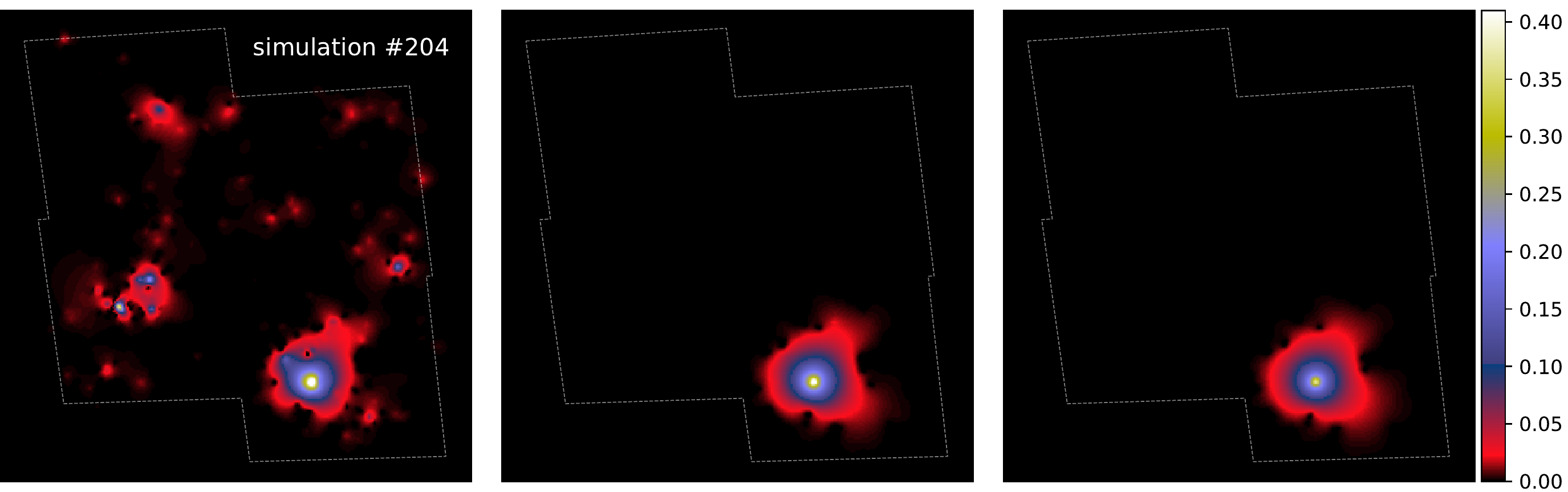}\\
\includegraphics[width=17.8cm]{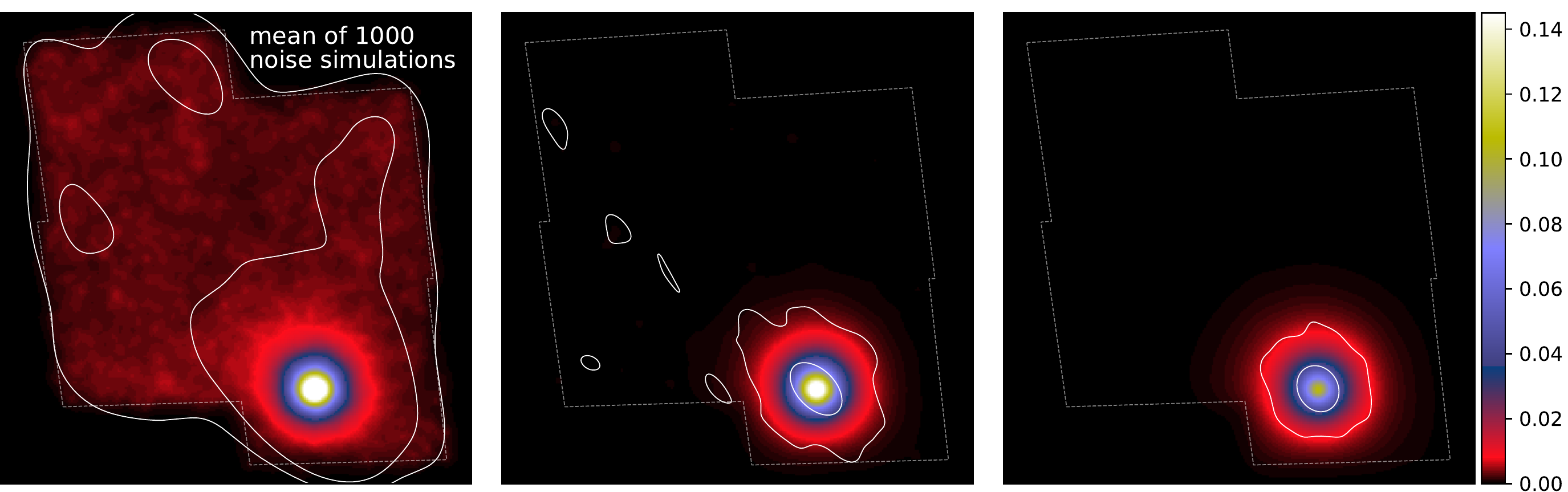}
\caption{Glimpse2D mass reconstructions of a simulated NFW halo within the HST/ACS footprint. Maps
         appearing in the left, middle, and right columns have been generated using the regularization 
         parameters $\lambda=2.0$, 3.0, and 4.0, respectively. \textit{Top row:} Shape noise--free
         mass maps reconstructed from the reduced shear of the NFW halo at $z=0.2$. The source galaxy
         positions are random with a number density matching the \cta{J14} catalog.
         \textit{Middle row:} Typical example of a noise simulation mass map. Higher $\lambda$ values
         effectively suppress the false peaks that appear for $\lambda=2.0$.
         \textit{Bottom row:} Mean of 1000 noise realizations. Contours represent the 1 and 2$\sigma$
         levels of distribution of highest amplitude peak positions.}
\label{fig:fig8}
\end{figure*}

We generate simulated data by injecting a Navarro-Frenk-White (NFW) \citep{NFW.1997} halo mass 
profile into the ACS footprint at $z=0.2$, the redshift of the A520 cluster. We place it at 
the position of P4 based on the original data reconstructions (see Figure \ref{fig:fig2}) with 
size and concentration parameters of $r_{200c}=0.85$ Mpc and $c=3.5$, respectively. This results 
in a mass $M_{200c}=4.9 \times 10^{13}~\mathrm{M}_\mathrm{sun}$, corresponding approximately to 
the average of the column masses derived for P4 in \cta{C12} and \cta{J14}. We compute the 
(reduced) shear field analytically \citep{WB.2000} and carry out Glimpse2D reconstructions 
without shape noise for $\lambda\in[1.0, 5.0]$ in steps of $\Delta\lambda=0.1$. Source galaxies 
are placed randomly within the field with a number density matching that of the \cta{J14} data.
The results for $\lambda=2.0$, 3.0, and 4.0 are shown in the top row of Figure \ref{fig:fig8}. 
The three maps are representative of all the reconstructions in that they are hardly 
distinguishable in terms of the morphology and position of the mass peak. The primary difference 
among the results shown is that the $\lambda=3.0$ and 4.0 peak amplitude is about 4\% lower than 
for $\lambda=2.0$. This example shows that Glimpse2D is able to accurately recover mass peaks 
in a noise-free setting even for relatively low values of $\lambda$. This is also consistent 
with the results of \citet{LSL.etal.2016}, in particular their experiment in Section 3.5.

We next test the impact of noise by assigning intrinsic shapes to the source galaxies drawn from
the \cta{J14} catalog. Our goal is to study the statistical behavior of Glimpse2D in simulations 
with a similar noise context to that of the original data. Note that using the galaxy 
ellipticities from the data this way does not perfectly capture the true noise properties for two reasons. 
The first is that the spatial variation of the galaxy density across a simulation field will not 
match that of the original data, and the second is that the shear produced by the A520 cluster is
still contained within the ellipticities. 
We do not expect the effect of the latter to be significant, as the shear remains weak throughout
most of the field and is further diluted by our random sampling of the catalog ellipticities.
Furthermore, because we are aiming merely for a guide to setting $\lambda$ in the real data 
reconstructions, the galaxy distribution need not match exactly---we can establish the 
expected behavior on average by carrying out a large number realizations of the noise where
the ellipticity distributions still match the data well. We have generated 1000 such noise 
simulations.

A typical example of a noise simulation with the same NFW profile described above is shown in
the second row of Figure \ref{fig:fig8}. From left to right, the maps represent Glimpse2D 
reconstructions for $\lambda=2.0$, 3.0, and 4.0. With $\lambda=2.0$, the input halo is clearly
recovered at the correct location; at this level, however, many spurious noise peaks appear as 
well. As discussed in Section \ref{sec:method}, this reflects the fact that lower $\lambda$ 
values are better at preserving smaller features but at the expense of signal to noise, i.e.,
our confidence that they are real. Looking closer at the region containing the true mass peak, 
we see that false peaks can arise close enough to the true peak that one might mistake them 
as extensions of it. Importantly, raising $\lambda$ to 3.0 (center panel) effectively suppresses 
this as well as all other noise peaks in the field. For this particular realization, raising 
$\lambda$ further does not improve the quality of the reconstruction in terms of eliminating 
false detections.

Finally, we examine the mean of the 1000 noise simulations for the same three $\lambda$
values as above, which are shown in the bottom row of Figure \ref{fig:fig8}. As expected, the
$\lambda=2.0$ map exhibits significantly more variation across the field due to noise compared 
to higher values. In all three cases, the true mass peak is smoothed out (and isotropized),
which we also expect due to small variations in the reconstructed peak amplitude and position
based on the particular noise realization. Overlaid are contours indicating the 1 and 2$\sigma$ 
levels of the distribution of the highest amplitude peak position in each map for a given 
$\lambda$. As such, the contours are not showing specifically the dispersion of the true
peak position, since the highest peak in a given map might be due to noise. This is apparent
in the $\lambda=2.0$ case, where clearly many false peaks appear with higher amplitude than
the true peak. Instead, this is a simple test to verify our expectation that the features
detected in higher $\lambda$ reconstructions can be interpreted as being part of the real 
signal. By $\lambda=4.0$, the contours have closed in on the true peak location. Note that in 
this case, since essentially all noise peaks are suppressed at this level, the contours 
indeed approximate the scatter in the recoverability of the true peak position.

The results described in this appendix motivate the range of $\lambda$ values used 
in the various mass reconstructions of the main text. Furthermore, we can take the size of 
the contours on the $\lambda=4.0$ map as an estimate of the typical variability of a peak. 
The 1 and 2$\sigma$ regions correspond to a radius of approximately 80 and 200 kpc, respectively, 
at the redshift of the cluster. The 150 kpc search radius about each peak in the bootstrap 
analysis of Section \ref{ssec:bootstrap} seems therefore reasonable.

\bibliographystyle{aasjournal}
\bibliography{refs}

\end{document}